\documentclass[epjST]{svjour}
\usepackage{graphics}
\usepackage{amsfonts,amsmath,xcolor,epsfig,epstopdf,multirow}
\newcommand{\SU}[1]{\ensuremath{\mathrm{SU}( #1 )}}
\newcommand{\Un}[1]{\ensuremath{\mathrm{U}( #1 )}}

\newcommand{\On}[1]{\ensuremath{\mathrm{O}( #1 )}}

\newcommand{\SpR}[1]{\ensuremath{\mathrm{Sp}( #1,\mathbb{R} )}}
\newcommand{\half}{\ensuremath{\textstyle{\frac{1}{2}}}}

\newcommand{\betb}{\begin{tabular}{p{4.0cm}p{9.0cm}}}
\newcommand{\entb}{\end{tabular}}

\newcommand{\hw}{\ensuremath{\hbar\Omega}}
\newcommand{\ph}[1]{\ensuremath{#1}p-\ensuremath{#1}h}

\newsavebox{\smlmat}% Box to store smallmatrix content
\savebox{\smlmat}{$\left(\begin{smallmatrix} {\bf A} & \hspace{6pt} &  {\bf B} \\  {\bf C} & \hspace{6pt}  & {\bf D} \end{smallmatrix}\right)$}
\begin{document}
\title{Emergent symplectic symmetry in atomic nuclei}
%Role of Symmetries in Nuclear Physics
\subtitle{\textit{Ab initio} symmetry-adapted no-core shell model}
\author{Kristina D Launey\inst{1}\fnmsep\thanks{\email{klauney@lsu.edu}} \and Tom\'a\v{s} Dytrych\inst{1,2}
\and Grigor H Sargsyan\inst{1} \and Robert B Baker\inst{3}
 \and Jerry P Draayer\inst{1}  
}
\institute{Department of Physics and Astronomy, Louisiana State University, Baton Rouge, LA 70803 
\and 
Nuclear Physics Institute, Academy of Sciences of the Czech Republic, 250 68 \u{R}e\u{z}, Czech Republic 
 \and 
Institute of Nuclear and Particle Physics, and Department of Physics and Astronomy, Ohio University, Athens, OH 45701}
\abstract{
Exact symmetry and symmetry-breaking phenomena play a key role in gaining a better understanding of the physics of many-particle systems, from quarks and atomic nuclei,  through molecules and galaxies.  In nuclei, exact and dominant symmetries  such as rotational invariance, parity, and charge independence have been clearly established. Beyond such symmetries, the nature of nuclear dynamics appears to exhibit a high degree of complexity, and only now, we show the fundamental role of an emergent approximate symmetry in nuclei, the symplectic \SpR{3} symmetry, as clearly unveiled  from {\it ab initio} studies that start from realistic interactions. In this article,  we detail and enhance our recent findings presented in Physical Review Letters 124 (2020) 042501, that  establish \SpR{3}  as a remarkably good symmetry of the strong interaction, and point to the predominance of a few equilibrium nuclear shapes (deformed or not) with associated vibrations and rotations that preserve the symplectic  \SpR{3}  symmetry. Specifically, we find that the structure of  nuclei below the calcium region in their ground state, as well as in their low-lying excited states and giant resonances, respects this symmetry at the 60-80\% level.  } %end of abstract
\maketitle
\section{Introduction}
\label{intro}
We have recently shown through first-principle large-scale nuclear structure calculations that the special nature of the strong nuclear force determines  highly regular patterns in nuclei that can be tied to an emergent approximate symmetry \cite{DytrychLDRWRBB20}. We find that this symmetry is remarkably ubiquitous, regardless of the type of the nucleus and the particular strong interaction heritage, and mathematically tracks with a symplectic group. For a set of $A$ particles, the symplectic symmetry \SpR{3} is based on a very basic concept: linear canonical transformations of particle coordinates and momenta that preserve the  fundamental Heisenberg commutation relation, as detailed in Sect. \ref{symm}. The most interesting insight, however, arises from a complementary perspective:  the symplectic symmetry has been recognized to preserve an equilibrium shape under transformations, such as rotations, orientations in space, and vibrations \cite{Rowe13}.
This, in turn, has important implications to our understanding  of the physics of nuclei: The approximate symplectic symmetry is manifested in nuclear states as the predominance  of only a few symplectic irreducible representations
%The approximate symmetry we observe in nuclei is manifested in nuclear states that are described by only a few symplectic irreducible representations 
(or irreps, subspaces of configurations that preserve the symmetry). 
Hence, we now understand that nuclei  are made of only a few equilibrium shapes, deformed or not, with associated vibrations and rotations. 

In this paper, we detail that  dominant features  of light to intermediate-mass nuclei (below the calcium region), including their low-lying excited states and giant resonances, track with the symplectic symmetry and naturally emerge from first-principle  (\textit{ab initio}) considerations, even in close-to-spherical nuclear states without any recognizable rotational properties. Therefore, the present outcomes not only explain but also predict the emergence of nuclear collectivity.

To study this without limitations within the interaction and approximations during the many-body nuclear simulations, we use the {\it ab initio} symmetry-adapted no-core shell model (SA-NCSM) \cite{DytrychSBDV_PRL07,DytrychLDRWRBB20,LauneyDD16}. The model  starts with realistic interactions tied to elementary particle physics considerations and fitted to nucleon-nucleon scattering data, and uses  symmetry-adapted (SA) bases based on the  \SpR{3} symmetry and its subgroup, the deformation-related \SU{3}. The \SpR{3} approximate symmetry is utilized to dramatically reduce computational resources required in \textit{ab initio} large-scale modeling of nuclear structure and reactions. This, in turn, pioneers predictions in open-shell deformed intermediate-mass nuclei, such as Ne and Mg isotopes \cite{DytrychLDRWRBB20,Henderson:2017dqc,Ruotsalainen19,PhysRevC.100.014322}, and targets  short-lived isotopes with enhanced deformation or cluster substructure along various nucleosynthesis pathways, especially where experimental measurements are incomplete or not available.

\section{Symplectic and SU(3) symmetries}
\label{symm}
It is well known that  SU(3)  \cite{Elliott58,Moshinsky62,DraayerSU3_1,MoshinskyPSW75,HechtZ79}  is the symmetry group of the spherical harmonic oscillator (HO) that underpins the 
shell model \cite{MayerJ55} and the valence-shell \SU{3} (Elliott) model  \cite{Elliott58,Elliott58b,ElliottH62} (for technical details of \SU{3}, see Ref. \cite{Kota20}).  
The Elliott  model has been shown  to naturally describe rotations of a deformed nucleus without the need for breaking rotational symmetry. 
The key role of deformation in nuclei and the coexistence of low-lying quantum states in a single nucleus characterized by configurations with different quadrupole moments 
\cite{HeydeW11} 
makes the quadrupole moment a dominant fundamental property of the nucleus. Hence, the quadrupole moment  and the monopole moment or ``size" of the nucleus, along with nuclear masses, establishes the energy scale of  the nuclear problem. Indeed, the nuclear monopole  and quadrupole moments underpin the essence of  symplectic \SpR{3} symmetry. 

The symplectic group \SpR{3} consists of all {\it particle-independent} linear canonical transformations of the single-particle phase-space observables, the positions $\vec r_i$ and momenta $\vec p_i$ (with particle index $i=1,\dots, A$ and spacial directions $\alpha,\beta=x,y,z$)
\begin{eqnarray}
r'_{i\alpha}&=&\sum_{\beta}A_{\alpha\beta}r_{i\beta}+B_{\alpha\beta}p_{i\beta}\\
p'_{i\alpha}&=&\sum_{\beta}C_{\alpha\beta}r_{i\beta}+D_{\alpha\beta}p_{i\beta}
\end{eqnarray}
 that preserve the Heisenberg commutation relations $[r_{i\alpha},p_{j\beta}]=i\hbar \delta_{ij}\delta_{\alpha \beta}$ \cite{Rowe85,Rowe13,LauneyDD16}. Generators of these transformations, symbolically denoted as matrices ${\bf A}$,  ${\bf B}$, ${\bf C}$, and ${\bf D}$, are constructed as ``quadratic coordinates" in phase space, $\vec r_i$ and  $\vec p_i$, and, most importantly, sum over all the particles and act on the space orientation  [on the contrary, the generators of the complementary \On{A} sum over the three spatial directions and act on the particle index, with a growing complexity with increasing particle number]. Hence, the generators include physically relevant operators: the total kinetic energy ($\frac{p^2}{2}=\frac{1}{2}\sum_{i}{\vec p_i \cdot \vec p_i}$), the monopole moment ($r^2=\sum_{i}{\vec r_i \cdot \vec r_i}$), the quadrupole moment ($Q_{2M}=\sqrt{16\pi/5 }\sum_ir_i^2Y_{2M}(\hat r_i)$), the orbital momentum ($\vec L=\sum_{i}\vec r_i \times \vec p_i$), and the many-body harmonic oscillator Hamiltonian ($H_0=\frac{p^2}{2}+\frac{r^2}{2}$).

Another key feature is that a single-particle \SpR{3} irrep spans \textit{all} positive-parity (or
negative-parity) states for a particle in a three-dimensional spherical or triaxial (deformed) harmonic oscillator.
Not surprisingly, the symplectic \SpR{3} symmetry, 
the underlying symmetry of the symplectic rotor model \cite{RosensteelR77,Rowe85}, 
has been found to play a key role across the nuclear chart -- from the lightest systems  \cite{RoweTW06,DreyfussLTDB13}, through intermediate-mass nuclei \cite{DraayerWR84,TobinFLDDB14,LauneyDD16}, up  to strongly deformed nuclei of the rare-earth and actinide regions \cite{Rowe85,CastanosHDR91,JarrioWR91,BahriR00}.
The results agree with experimental evidence that supports formation of enhanced deformation  and clusters in nuclei, as well as  vibrational and rotational patterns, as suggested by energy spectra, electric monopole and  quadrupole transitions, radii and quadrupole moments \cite{HeydeW11,Henderson:2017dqc,FreerHKLM18}. 
And while these earlier algebraic models have been very successful in explaining dominant nuclear patterns, they have assumed symmetry-based approximations and have often neglected symmetry mixing. This establishes \SpR{3} as an \textit{effective 
symmetry}\footnote{
A familiar example for an effective symmetry is \SU{3}. While the Elliott model with a single \SU{3} irrep explains ground-state rotational states in deformed nuclei, the \SU{3} symmetry is, in general, largely mixed, mainly due to the spin-orbit interaction (nonetheless, \SU{3} has been shown to be an excellent quasi-dynamical symmetry, that is,   each rotational state has almost the same \SU{3} content \cite{BahriR00}).
} 
for nuclei, which may or may not be badly broken in realistic calculations. It is then imperative to probe if this symmetry naturally arises within an \textit{ab initio} framework, which will, in turn, establish its fundamental role.

\section{Many-body symmetry-adapted (SA) framework}
\label{sec:1}
{\it Ab initio} approaches (e.g., see Ref. \cite{Launey16reviewbook} and references therein)
build upon a ``first principles" foundation, namely, the properties of only two or three nucleons that are often tied to symmetries and symmetry-breaking patterns of the underlying quantum chromodynamics theory.
We utilize the {\it ab initio} nuclear shell-model theory \cite{CaurierMNPZ05,NavratilVB00,BarrettNV13} that solves the many-body Schr\"odinger equation for $A$ particles,
\begin{equation}
H \Psi(\vec r_1, \vec r_2, \ldots, \vec r_A) = E \Psi(\vec r_1, \vec r_2, \ldots, \vec r_A).
\label{ShrEqn}
\end{equation}
 In its most general form, it is an exact many-body ``configuration interaction" method, for which the interaction and basis configurations are as follows. 
\newline
{\em Interaction. } 
The intrinsic non-relativistic nuclear  Hamiltonian $H = T_{\rm rel} + V_{NN}  + V_{3N} + \ldots + V_{\rm Coulomb}$  includes
the relative kinetic energy $T_{\rm rel} =\frac{1}{A}\sum_{i<j}^A\frac{(\vec p_i - \vec p_j)^2}{2m}$ ($m$ is the nucleon mass),  
the nucleon-nucleon $V_{NN}$ and, possibly, three-nucleon $V_{3N}$ interactions and beyond, typically derived in the chiral effective field theory \cite{BedaqueVKolck02,EpelbaumNGKMW02,EntemM03,Epelbaum06},
along with the Coulomb interaction between the protons. 
\newline
{\em Basis configurations.} 
The method adopts a complete orthonormal many-particle basis $\psi_k$, such as the antisymmetrized products of  single-particle states  of a spherical harmonic oscillator of  characteristic length $b=\sqrt{\hbar \over m\Omega}$. 
The expansion $\Psi(\vec r_1, \vec r_2, \ldots, \vec r_A) = \sum_{k} c_k \psi_k(\vec r_1, \vec r_2, \ldots, \vec r_A)$
 renders Eq. (\ref{ShrEqn}) into a matrix eigenvalue equation with unknowns $c_k$,
$
\sum_{k'} H_{k k'} c_{k'} = E c_k,
$
where the many-particle Hamiltonian matrix elements
$H_{k k'} = \langle \psi_k | H | \psi_{k'} \rangle$ are calculated for the given interaction and the solution $\{c_k^2\}$ defines a set of probability amplitudes. 

However, in {\it ab initio} shell-model calculations the complexity of the nuclear problem dramatically increases with the number of particles, and when expressed in terms of literally billions of  shell-model basis states, the structure of a nuclear state is unrecognizable.
But expressing it in a more informative basis, the
symmetry-adapted (SA) collective basis \cite{DytrychSBDV_PRL07,LauneyDD16}, leads to a major breakthrough: we observe the incredible simplicity of nuclear low-lying states  and the dominance  of an approximate  symmetry of nuclear dynamics, the symplectic \SpR{3} symmetry, which  together with its slight symmetry breaking  naturally describe atomic nuclei.

The SA-NCSM is reviewed in Ref. \cite{LauneyDD16} and has been first applied to nuclei below the calcium region using the \SU{3}-adapted basis \cite{DytrychLMCDVL_PRL12} and the \SpR{3}-adapted basis \cite{DytrychLDRWRBB20}. Briefly, the  many-nucleon basis states of the SA-NCSM are constructed using efficient group-theoretical algorithms \cite{DraayerLPL89} and are labeled according to  \SU{3}$_{(\lambda\,\mu)}\times$\SU{2}$_S$ by the total  intrinsic spin $S$ and $(\lambda\,\mu)$ quantum numbers with $\lambda=N_z-N_x$ and $\mu=N_x-N_y$, where $N_x+N_y+N_z=N\le N_{\rm max}$ for a total of $N$ HO quanta distributed in the $x$, $y$, and $z$ direction. Hence, e.g., $N_x=N_y=N_z$, or equally $(\lambda\,\mu)=(0 0)$, describes a spherical configuration, while $N_z$ larger than  $N_x=N_y$, or $\mu=0$, indicates prolate deformation. Hence, the model space is reorganized to subspaces that have fixed deformation, specified by \Un{3} and its quantum numbers $N(\lambda\,\mu)$. One can further organize these  deformed configurations to subspaces that are associated with a fixed shape (here referred to as ``equilibrium shape" or simply ``shape"), labeled by a single deformation $N_0(\lambda_0\,\mu_0)$. These subspaces, specified by \SpR{3},  include the equilibrium shape, its vibrations (referred to as ``dynamical shapes")  and rotations. E.g., a symplectic irrep $0(8\,0)$ consists of a prolate equilibrium shape with $\lambda_0=8$ and $\mu_0=0$ in the \ph{0} (0-particle-0-hole) subspace (valence shell), along with many other deformed configurations, or dynamical shapes, that include particle-hole excitations to higher shells.

As for the inter-nucleon interaction, it is unitarily transformed to the SA basis, with matrix elements reduced with respect to \SU{3}$\times$\SU{2} (outlined in Refs. \cite{LauneyDDSD15} and  \cite{1937-1632_2019_0_183}). 
We note that while the model utilizes symmetry groups to construct the basis, calculations are not limited {\it a priori} by any symmetry and employ a large set of basis states that can,  if the nuclear Hamiltonian demands, describe a significant symmetry breaking.

In Refs. \cite{DytrychLMCDVL_PRL12,DytrychMLDVCLCS11,DytrychHLDMVLO14}, we have shown that the SA-NCSM can use a significantly reduced number of \SU{3}-adapted basis states (selected model spaces) as compared to the corresponding  large  complete $N_{\rm max}$  model space without compromising the accuracy for various observables, including energies, point-particle proton and matter rms radii, electric quadrupole and magnetic dipole moments, reduced electromagnetic B(E2) transition strengths, and electron scattering form factors.  In addition,
we have shown that, for these selected spaces, the size of the model space and the number of nonzero Hamiltonian matrix elements grow slowly with $N_{\rm max}$ \cite{DytrychMLDVCLCS11}, allowing the SA-NCSM to accommodate model spaces of larger $N_{\rm max}$ and to reach heavier nuclei, with recent SA-NCSM results  reported for $^{32}$Ne and $^{48}$Ti \cite{LauneySOTANCP42018}. 
\begin{figure}[th]
\begin{tabular}{ccc}
\includegraphics[width=0.28\textwidth]{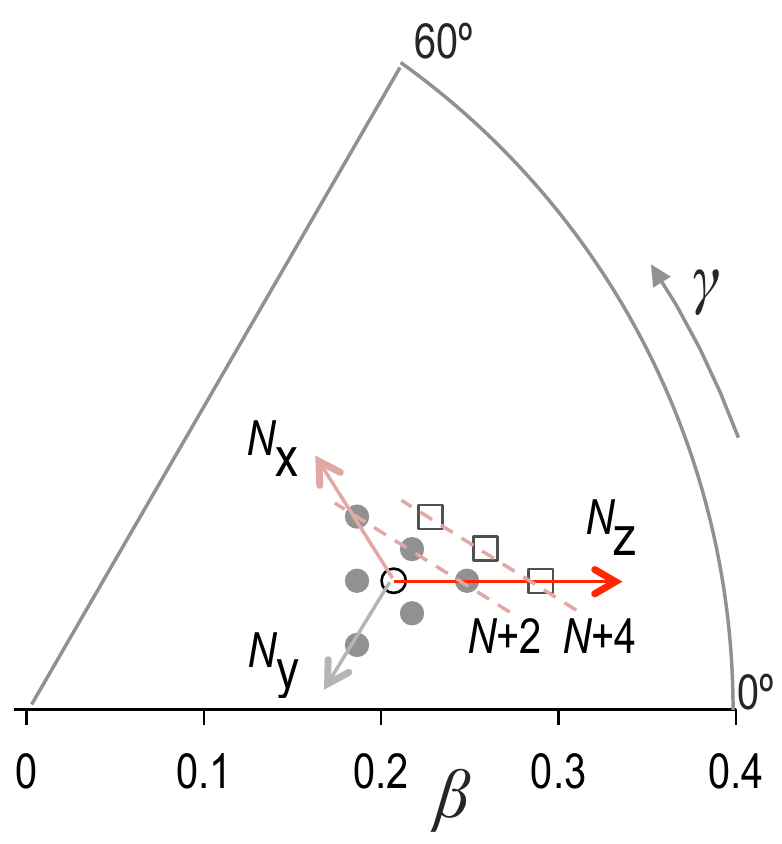} &
\includegraphics[width=0.5\textwidth]{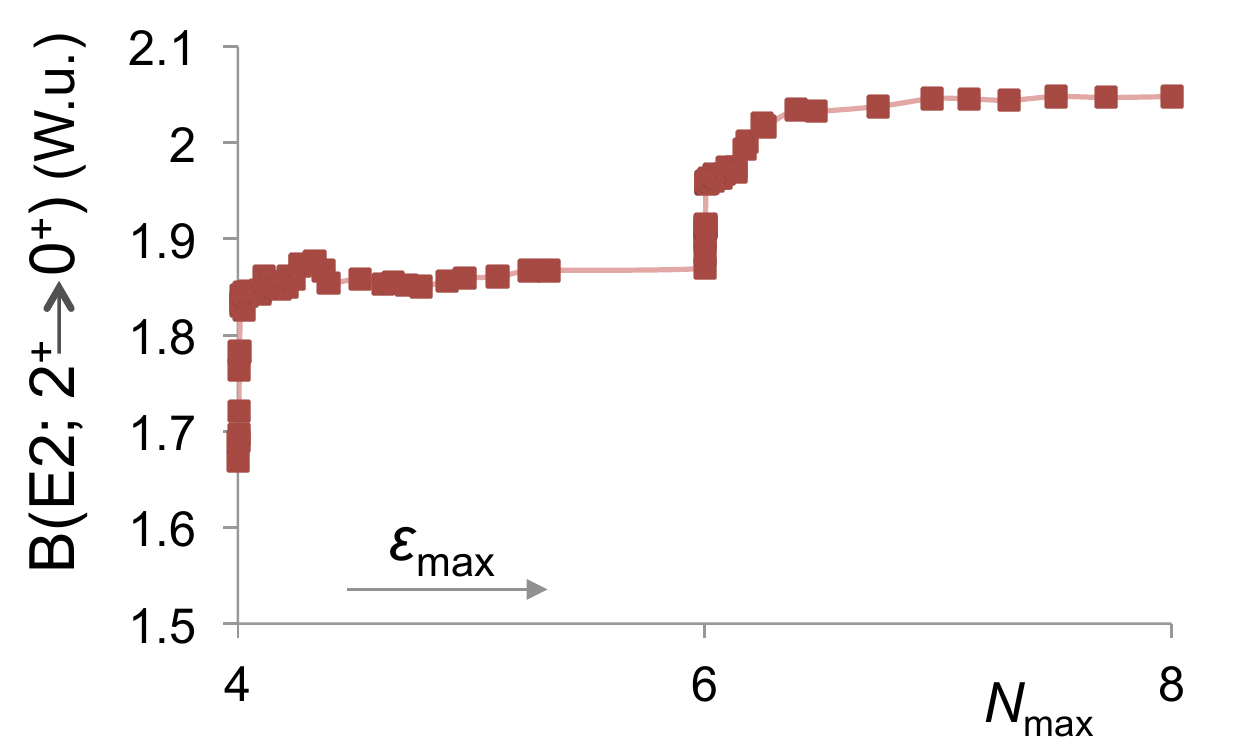} \\
(a) & (b) 
\end{tabular}
\caption{
(a) Selection of configurations in $N+4$ (open squares) based on dominant $N+2$ configurations, shown in terms of the deformation $\beta$ and triaxiality $\gamma$ shape parameters \cite{BohrMottelson69,CastanosDL88}.  $N+2$  configurations (filled circles) within a symplectic irrep are determined by the kinetic energy or quadrupole operator (for monopole or quadrupole symplectic excitations, respectively), or equivalently, by the number of all possible ways  two HO quanta  are distributed in the $x$, $y$, and $z$ directions.  (b) Convergence of the B(E2$;2_1^+ \rightarrow 0^+_{\rm gs})$ strength  in $^{12}$C  with the SA-NCSM selection cutoff, $\varepsilon_{\rm max}$, for the JISP16 $NN$ interaction  \cite{ShirokovMZVW07} and \hw=20 MeV. 
}
\label{C12vertical_convergence}
\end{figure}

To achieve this, in the SA-NCSM all basis states are kept up to a given $N$, while for higher $N$, the model space is down selected in a systematic way using \SpR{3} considerations (Fig. \ref{C12vertical_convergence}a). Configurations that are highly favored in the $N$ model space inform important configurations in the $N+2$ model space, which in turn inform the $N+4$ model space, etc., and those track with larger deformation $\beta$ along the $N_z$ axis (consistent with results in Refs. \cite{TobinFLDDB14,DreyfussLTDBDB16}). Notably, these $N+4$ configurations can be readily reached from the $N+2$ configurations in the $N_z$-$N_x$ plane by two excitations in the $z$ direction.  Hence, we can introduce a selection cutoff $\varepsilon_{\rm max}$, that is given by the fraction of the SA model space used. The order in which basis states are included is determined according to a weight $w(N_x,N_y,N_z+2)=\frac{P(N_x,N_y,N_z)}{\dim(N_x,N_y,N_z+2)}$, where $P(N_x,N_y,N_z)$ is the probability amplitude obtained in SA-NCSM calculations in the $N$ model space, and $\dim$ denotes the dimensionality of the configuration to be selected (spin degrees are omitted for simplicity). The prescription is then applied to $N+4$ up through $N_{\rm max}$. Similar to NCSM, a measure of convergence of the results is the degree to which the SA-NCSM obtains results independent of the model parameters $N_{\rm max}$, \hw, and $\varepsilon_{\rm max}$ [see Fig. \ref{C12vertical_convergence}(b) for increasing $\varepsilon_{\rm max}$]. Remarkably, even for small $\varepsilon_{\rm max}$ cutoffs, which correspond to drastically reduced model spaces, observables such as, e.g., B(E2) values are quite close to the converged results, a feature that further improves with $N_{\rm max}$. 
 A major advantage of the SA-NCSM is that  the SA model space can be down-selected to a subset of SA basis states that describe equilibrium and dynamical shapes, and within this selected model space the spurious center-of-mass motion can be factored out exactly~\cite{Verhaar60,Hecht71}.

In SA-NCSM calculations that use the \SpR{3}-adapted basis \cite{DytrychLDRWRBB20}, the basis is built from the \SU{3}-adapted basis. The difficulty stems from the fact that there are no known \SpR{3} coupling/recoupling coefficients, and one has to resort to innovative techniques. In our method, we adopt the \SU{3} scalar operator $\{A^{(2\,0)}\times B^{(0\,2)}\}^{(0\,0)}_{L=0 M=0}$ constructed by symplectic \SpR{3} generators, where $B^{(0\,2)}_{LM}=(-)^{L-M}(A^{(2\,0)}_{L-M})^\dagger$ is conjugate to $A^{(2\,0)}$ and moves a particle two shells down. This \SU{3} scalar operator is computed for a given set of basis states with the same $N(\lambda\,\mu)$; eigenvectors of this matrix realize \SpR{3}-adapted basis states and provide a unitary transformation from the \SU{3}-adapted basis to the \SpR{3}-adapted basis, while the known eigenvalues are used to assign each eigenvector to a specific symplectic irrep. Given the unitary transformation, the \SU{3}-decomposed  Hamiltonian is straightforwardly constructed for the new \SpR{3} basis. In selected model spaces, the resulting Hamiltonian matrix is then drastically small in size and its eigensolutions, the nuclear energies and states, can be  calculated without the need for supercomputers.
\begin{figure}[th]
\begin{tabular}{lll}
\includegraphics[width=0.30\textwidth]{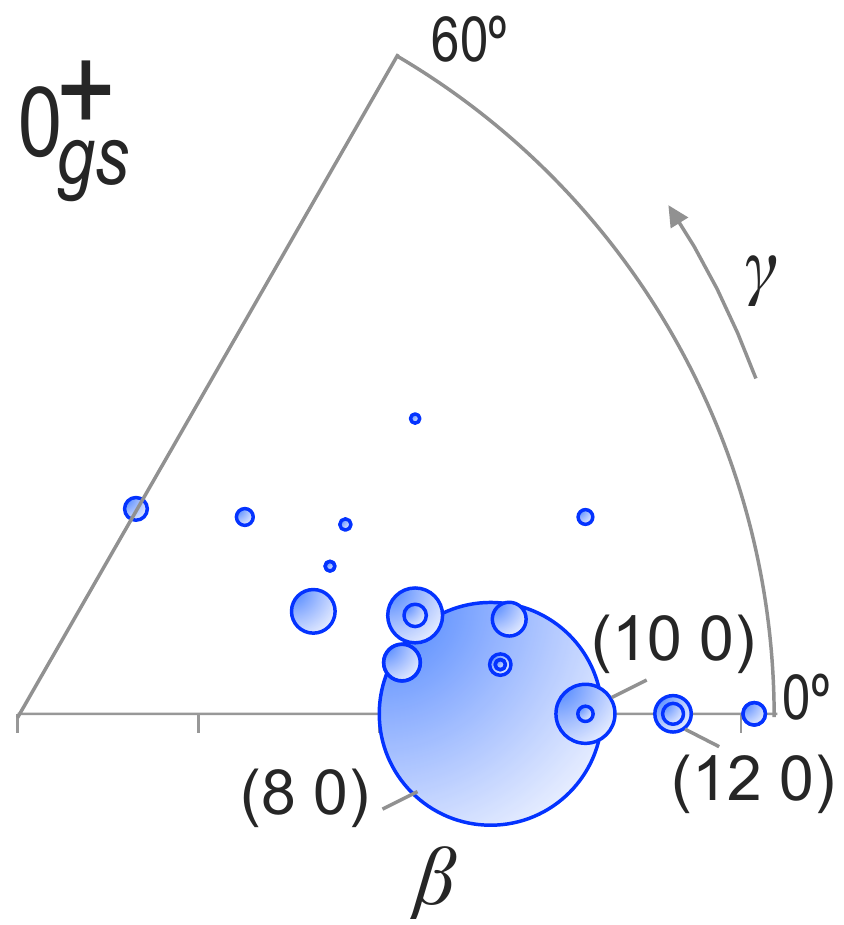} & 
\includegraphics[width=0.44\textwidth]{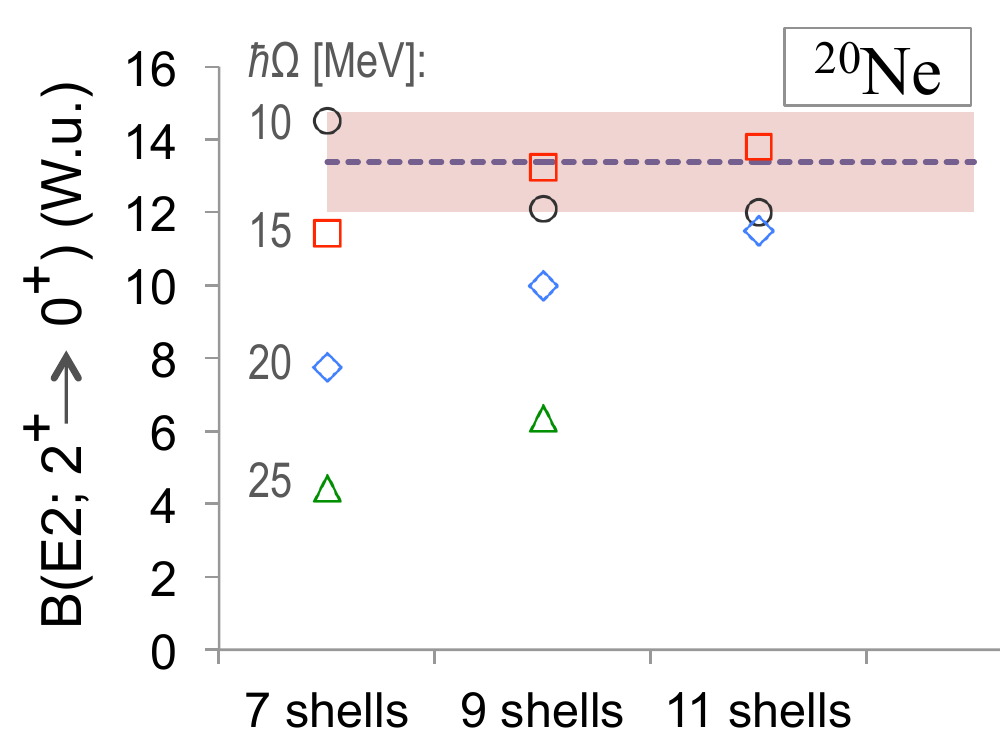}\\
(a) & (c) \\
\includegraphics[width=0.30\textwidth]{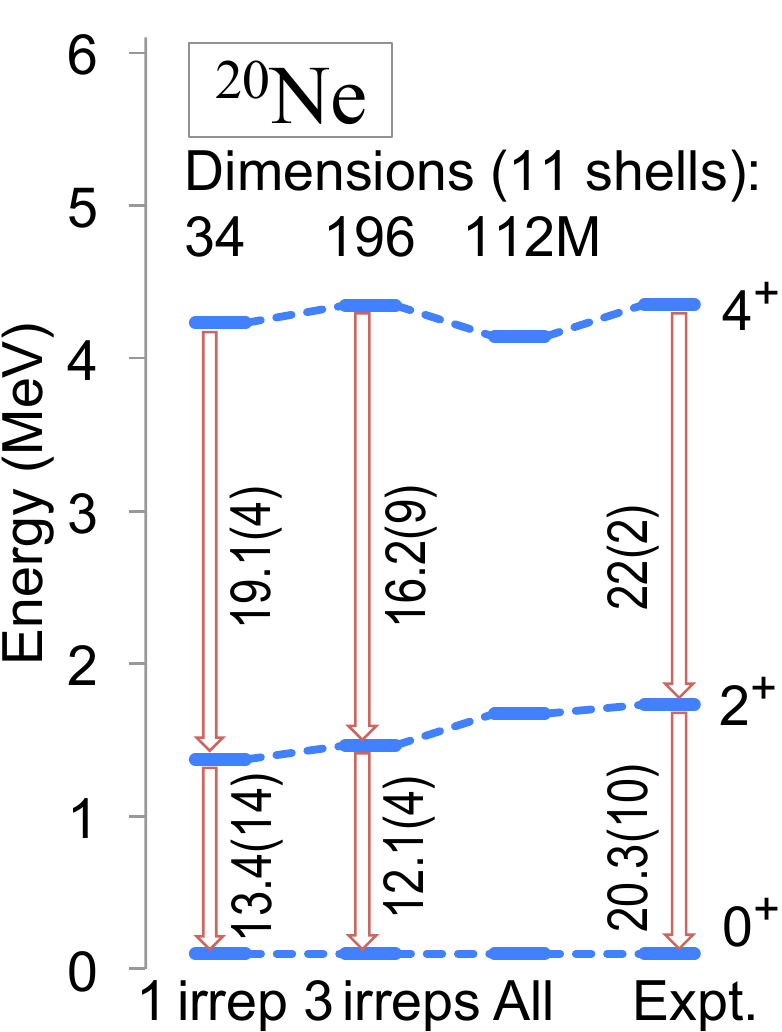}&
\includegraphics[width=0.44\textwidth]{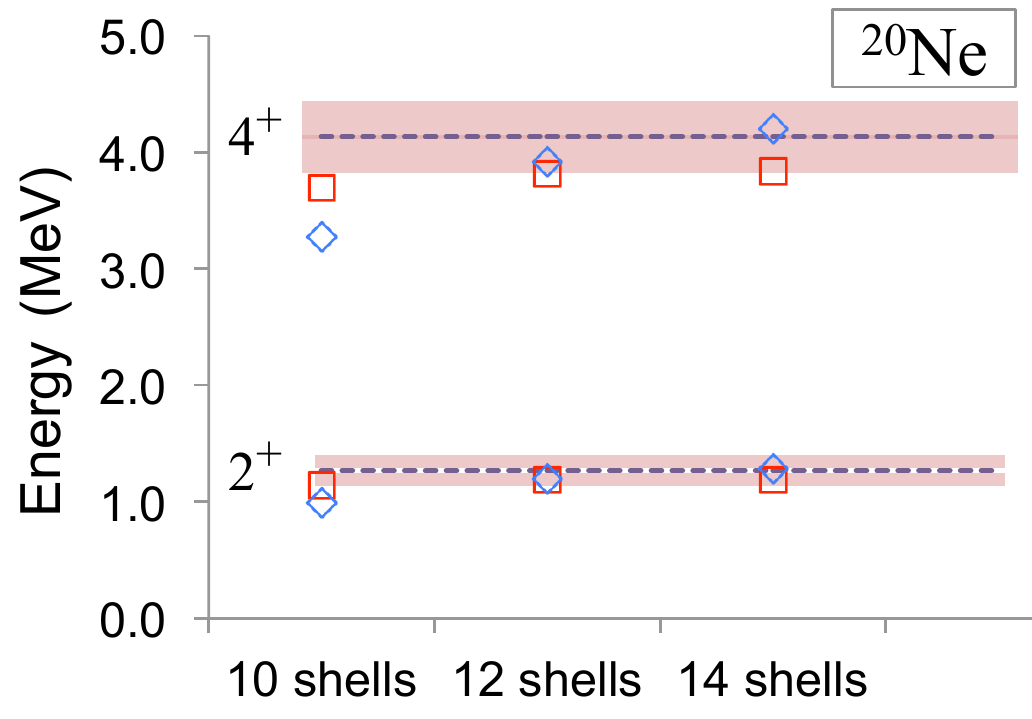}\\
(b) & (d) \\
\end{tabular}
\caption{
{\it Ab initio} SA-NCSM of $^{20}$Ne calculated   in the model space of 11 HO major shells with the NNLO$_{\rm opt}$ interaction \cite{Ekstrom13} and $\hw=15$ MeV:  (a) Contribution of the symplectic  \SpR{3} irreps, given by the size of the circles,  that make up the ground state of $^{20}$Ne, calculated  with an \SU{3}-adapted basis in a selected model space of $112\times10^6$ basis states (labeled as ``All");  each irrep is specified by its equilibrium shape, labeled by the shape deformation $\beta$ and triaxiality $\gamma$, with the largest contribution coming from the $(8\,0)$ shape.
(b) Observables calculated with the \SpR{3}-adapted basis  using only the  most dominant 1 or 3 symplectic irreps, as compared to experiment (``Expt."); dimensions of the largest model spaces used  are also shown. 
Energies and reduced B(E2) transition strengths (in W.u.) are reported  for extrapolations to infinitely many  shells of converging results  across variations in the number of shells and \hw. (c)-(d) SA-NCSM calculations and the extrapolated central value (dashed line) and uncertainties (red shaded band) for the single most deformed \SpR{3} irrep used in (b).
}
\label{sp_en_Ne20}      
\end{figure}

%%%%%%%%%%%%%%%%
%%%%%%%%%%%%%%%%
\section{Nature's preference: Approximate symplectic symmetry from first principles}
We report on the remarkable outcome, as unveiled from first-principle calculations below the calcium region, that nuclei exhibit relatively simple physics. We now understand that a low-lying nuclear state is predominantly composed  of a few equilibrium shapes that vibrate and rotate, with each shape 
characterized by a single symplectic irrep. E.g., in $^{20}$Ne, there is a single most predominant irrep, $(8\,0)$, that make up about 70\% or more of the ground state (Fig. \ref{sp_en_Ne20}a) and its rotational excitations \cite{DytrychLDRWRBB20}. Using this single \SpR{3} irrep,  it is notable that even excitation energies and B(E2) strengths fall closely to  the experimental data (Fig. \ref{sp_en_Ne20}b). Indeed, 
E2 transitions are determined by the quadrupole operator $Q$, an  \SpR{3} generator that does not mix symplectic irreps -- hence, the largest fraction of these transitions, 
and hence nuclear collectivity,  necessarily emerges within this most dominant symplectic irrep (similarly for rms radii, since $r^2$ is also a  symplectic generator). 

To report observables, we use extrapolations to the infinite number of shells (independent of model parameters $N_{\rm max}$ and \hw). They are based on the fast convergence we observe for nuclear properties, such as energies and B(E2) strengths, within a set of symplectic irreps and  around an optimum \hw~value (Fig. \ref{sp_en_Ne20}c \& d).  Hence, for data on a converging trend, one can use the Shanks transformation ansatz for a quantity $X_{\infty}=\sum_{n=0}^{\infty}x_n$ such that  $X_{N}=\sum_{n=0}^{N}x_n$ is given by  $X_{N}=X_\infty +Aq^N$ for large $N$, where $0<q<1$ \cite{Shanks55,BenderO78}. $X_ \infty$ is the extrapolated result independent of the basis parameters (or the ``full-space" result  within the set of symplectic irreps). The truncation error at each order $N$ of the series expansion is given by $\mathcal O (q^N)$; the leading-order error $x_{N+1}=X_{N+1}-X_N$ is thus $\mathcal O (qx_{N})$, which is consistent with effective-field-theory expansions  for a sufficiently small expansion parameter (see, e.g., \cite{PhysRevC.96.024003}). For results that largely depend on \hw, extrapolated values for each \hw~significantly deviate; however, for \hw~ values around the optimum one, deviations in $X_\infty$ are drastically reduced, leading to relatively small uncertainties in the quoted extrapolated values.

\subsection{Symmetry in low-lying excited states and giant resonances}
The near symmetry is not restricted to ground states, but extends to  low-lying states. E.g., some yrast states have almost identical \SpR{3} structure to that of the ground state (e.g., see Ref. \cite{DytrychLDRWRBB20} for $3_1^+$ and $2_1^+$ in $^6$Li, and for $2_1^+$ and $4_1^+$ in $^{20}$Ne; see also Fig. \ref{su3_12C} for $^{12}$C).  Practically the same symplectic content observed  is a rigorous signature of rotations of a shape and can be used to identify members of a rotational band and enhanced B(E2) strengths. 
\begin{figure}[th]
\begin{tabular}{cc}
\includegraphics[width=0.5\textwidth]{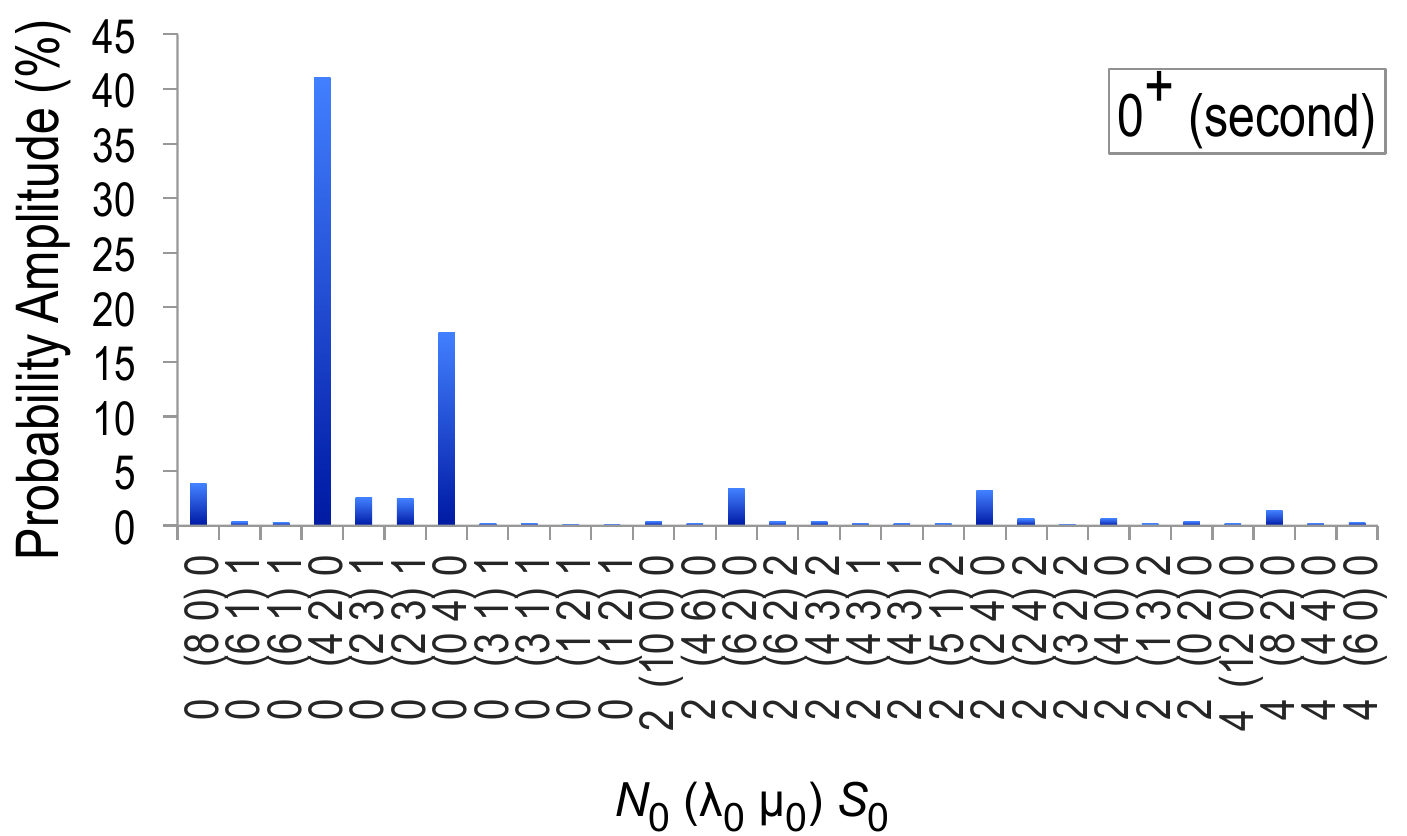} & \hspace{-13pt}
\includegraphics[width=.5\textwidth]{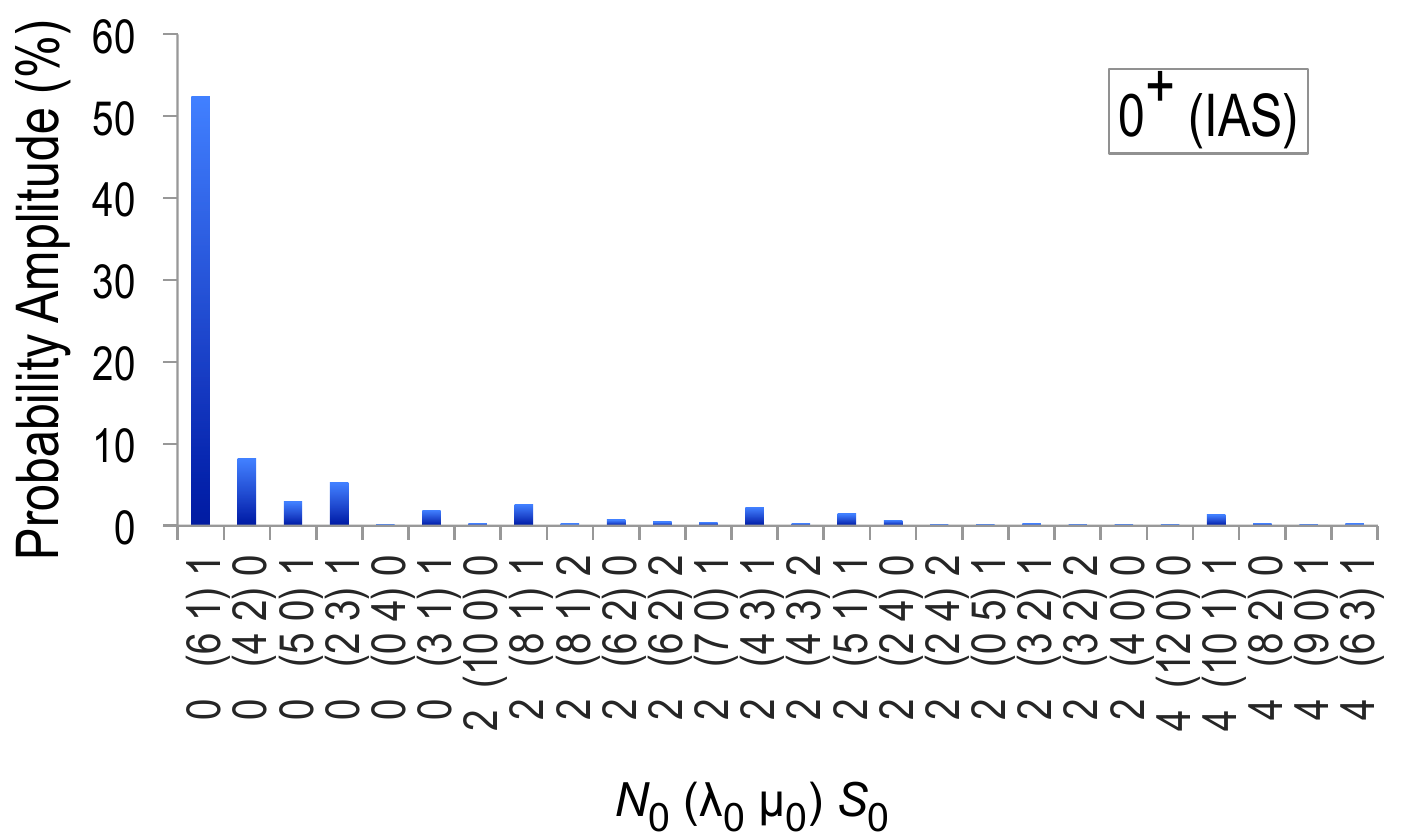} \\
{\footnotesize (a)} & {\footnotesize(b) }
\end{tabular}
\caption{
Symplectic  \SpR{3} irreps  that make up (a) the second $0^+$  state and (b) the  lowest $0^+$  isobaric analog state (IAS) in $^{20}$Ne, labeled by the total HO excitations $N_0$,  \SU{3} labels $(\lambda_0\,\mu_0)$, and total intrinsic spin $S_0$ of the equilibrium shape.  
Results are reported for {\it ab initio} SA-NCSM calculations with the NNLO$_{\rm opt}$ interaction, for an \SU{3}-adapted basis  in a selected model space of 11 HO major shells and $\hw= 15$ MeV.
}
\label{sp_20Ne}      
\end{figure}

Furthermore, the lowest $1^-$ state in $^{20}$Ne is found to be largely dominated by the prolate $(9\, 0)S=0$ shape, with some  contribution from $(5\, 2)S=0$  \cite{DreyfussLESBDD20}. The second $0^+$ state in  $^{20}$Ne is dominated by two equilibrium shapes (Fig. \ref{sp_20Ne}a). It is interesting to note that the most deformed equilibrium shape for protons  in the valence shell is $(4\,0)$, and the same for neutrons, resulting in overall shapes of $(8\,0)$, $(4\,2)$, and $(0\,4)$, the first of which dominates the ground state (Fig. \ref{sp_en_Ne20}a), whereas the $(4\,2)$ and $(0\,4)$  shapes dominate the next $0^+$ state for the NNLO$_{\rm opt}$ interaction  (Fig. \ref{sp_20Ne}a). Another interesting $0^+$ state  in $^{20}$Ne  is the lowest isobaric analog state (IAS), which corresponds to the lowest $0^+$ state in the neighboring  $^{20}$Na and  $^{20}$F  isotopes (Fig. \ref{sp_20Ne}b). This state manifests a dominance of a single prolate shape $(6\,1)$ that is slightly less deformed as compared to the $(8\,0)$ shape and is the main shape of the  $^{20}$Na and  $^{20}$F lowest $0^+$ state where $(8\,0)$ is Pauli forbidden. It is important to emphasize that, besides the predominant irrep(s), there is a manageable number of symplectic irreps, each of which contributes at a level that is typically at least an order of magnitude  smaller.
\begin{figure}[th]
\begin{tabular}{cc}
\includegraphics[width=0.45\textwidth]{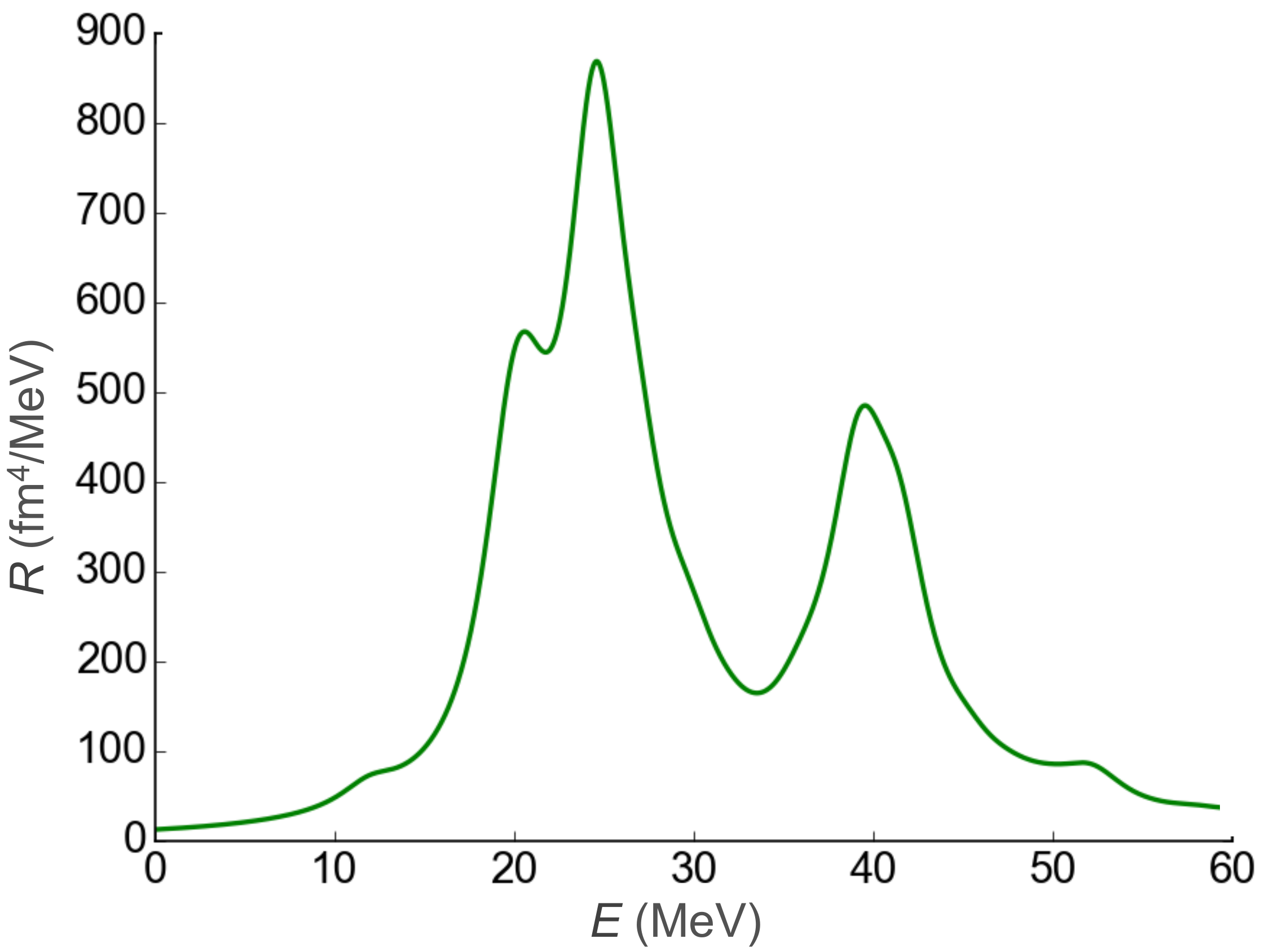} & \hspace{-13pt}
\includegraphics[width=.54\textwidth]{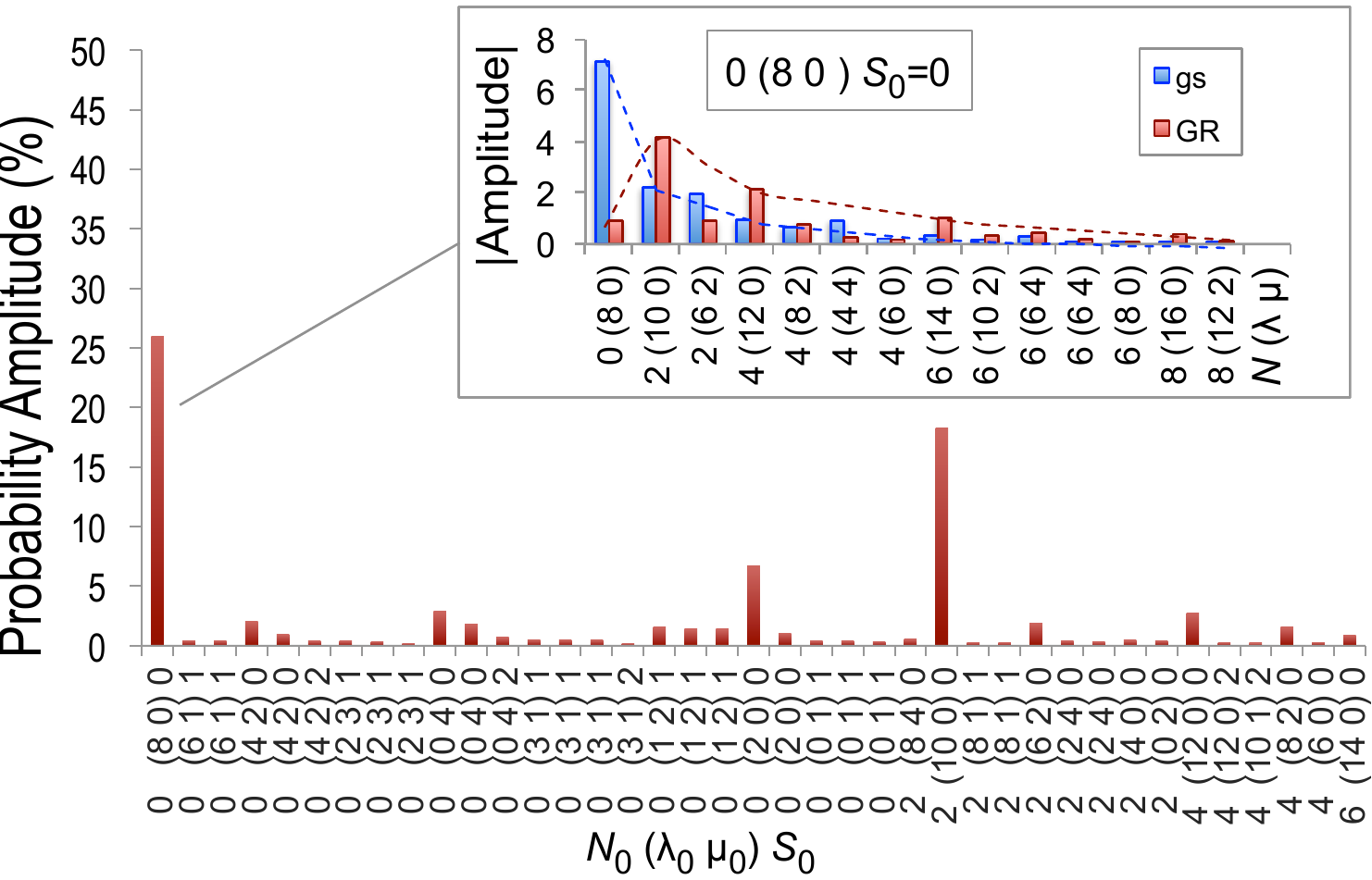} \\
{\footnotesize (a)} & {\footnotesize(b) }
\end{tabular}
\caption{
(a) Monopole response function for the $^{20}$Ne ground state vs. excitation energy, for a width of $\Gamma=2$ MeV of the Lorentzian kernel \cite{BakerLBND20}. (b) Symplectic  \SpR{3} irreps  that make up the $0^+$  excited state (``GR") that contributes the most to the giant resonance of $^{20}$Ne, labeled by the total HO excitations $N_0$,  \SU{3} labels $(\lambda_0\,\mu_0)$, and total intrinsic spin $S_0$ of the equilibrium shape.  Inset: Distribution of the $N_0=0(8\,0)S_0=0$ shape in this ``GR" state along \SU{3} configurations, labeled by the total HO excitations $N$ and \SU{3} labels $(\lambda\,\mu)$, as compared to the distribution of the same shape in the ground state (``gs").
Results are reported for {\it ab initio} SA-NCSM calculations that use the NNLO$_{\rm opt}$ interaction for $\hw= 15$ MeV, and with an \SU{3}-adapted basis  in a selected model space of (a) 13 and (b) 11 HO major shells. 
}
\label{sp_20NeGR}      
\end{figure}

Nuclear saturation properties can be informed by  nuclear breathing modes, or giant monopole resonances \cite{BakerThesis19}. To study these, we calculate the  response of the $^{20}$Ne ground state to an isoscalar monopole  probe $M_0=\half \sum_i r_i^2$  (Fig. \ref{sp_20NeGR}a). Since  the $M_0$ operator is a symplectic generator and does not mix symplectic irreps, the monopole response tracks the contribution of the  $(8\,0)$ shape, the predominant shape of  the $^{20}$Ne ground state, to all excited $0^+$ states. It is not surprising then that the distribution and the peak of the response function are consistent with the results of 
Ref. \cite{DytrychLDRWRBB20}, where the set of excited $0^+$ states with nonnegligible contribution of the \ph{1} vibrations of the ground-state shape $(8\,0)$ has been suggested to describe a fragmented giant monopole resonance with a centroid around $29$ MeV  and a typical wavefunction spread out to higher deformation due to vibrations \cite{BahriDCR90}, as compared to the ground state (Fig. \ref{sp_20NeGR}b, inset). Indeed, by examining the state that is related to the  peak in the response function, we find that two equilibrium shapes dominate (Fig. \ref{sp_20NeGR}b): $N_0=0(8\,0)S_0=0$ and $N_0=2(10\,0)S_0=0$. The $(8\,0)$ also dominates the ground state, where it peaks at $N=0$, while in the giant resonance state this shape peaks at the $N=2$ $(10\,0)S=0$ vibration (Fig. \ref{sp_20NeGR}b, inset).
Note that $N_0=2(10\,0)S_0=0$ is an equilibrium shape and $N=2$ $(10\,0)S=0$ is a dynamical shape, a vibration of the  $N_0=0(8\,0)S_0=0$ equilibrium shape, but  remarkably both have the same  \SU{3} quantum numbers.

\begin{figure}[th]
\begin{tabular}{cc}
\includegraphics[width=0.5\textwidth]{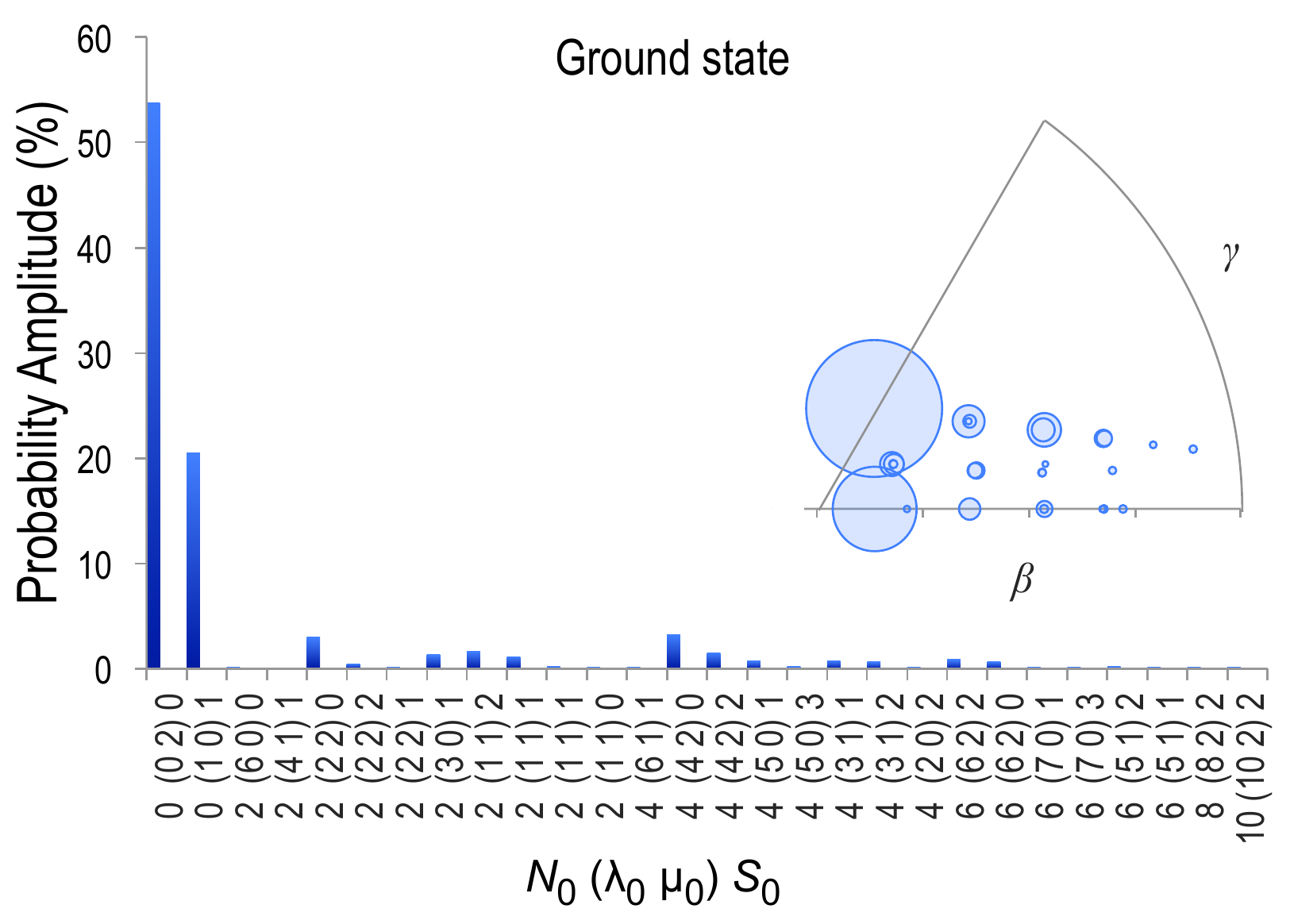} & \hspace{-13pt}
\includegraphics[width=0.5\textwidth]{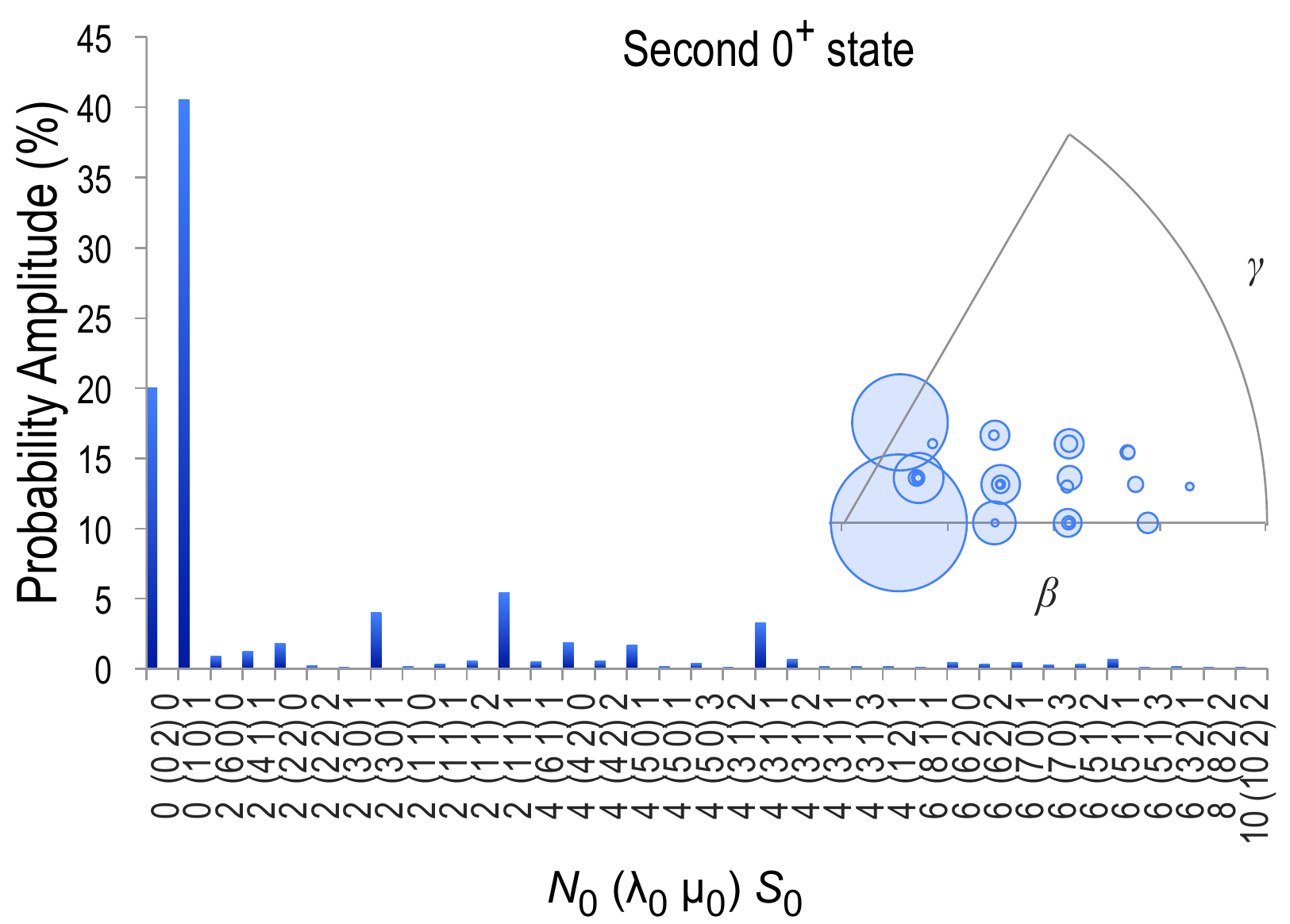} \\
{\footnotesize (a)} & {\footnotesize(b) }
\end{tabular}
\caption{
Symplectic  \SpR{3} irreps  that make up (a) the $0^+$ ground state and (b) the second $0^+$  state of $^{8}$He, labeled by the total HO excitations $N_0$,  \SU{3} labels $(\lambda_0\,\mu_0)$, and total intrinsic spin $S_0$ of the equilibrium shape, together with the corresponding  $\beta$-$\gamma$ plot.  
Results are reported for {\it ab initio} SA-NCSM calculations with the NNLO$_{\rm opt}$ interaction, for an \SU{3}-adapted basis  in a complete model space of 14 HO major shells and $\hw= 20$ MeV.
}
\label{sp_8He}      
\end{figure}

Similarly, the lowest two $0^+$ states in $^8$He, which has been suggested to be a halo nucleus, are made of a predominant shape  that contributes at the 40-55\% level and a secondary in importance shape with about 20\% contribution (Fig. \ref{sp_8He}). It is interesting to note that both shapes are ``opposite" in their deformation,  $(1\,0)$  is prolate  and the other one $(0\,2)$ is oblate (we note that another common convention associates  a positive  $\beta$ value with a prolate shape, whereas a negative $\beta$ indicates an oblate shape). While, in general,  $^8$He is considered to be spherical, the present outcome points to an interplay of two shapes in the ground state, which on average may appear to have a zero deformation, but with a B(E2) strength from its $2^+$ rotational state that constructively adds the nonzero contributions of both shapes. 

%%%%%%%%%%%%%%

\subsection{Sensitivity to the NN interaction}
\begin{figure}[th]
\includegraphics[width=0.95\textwidth]{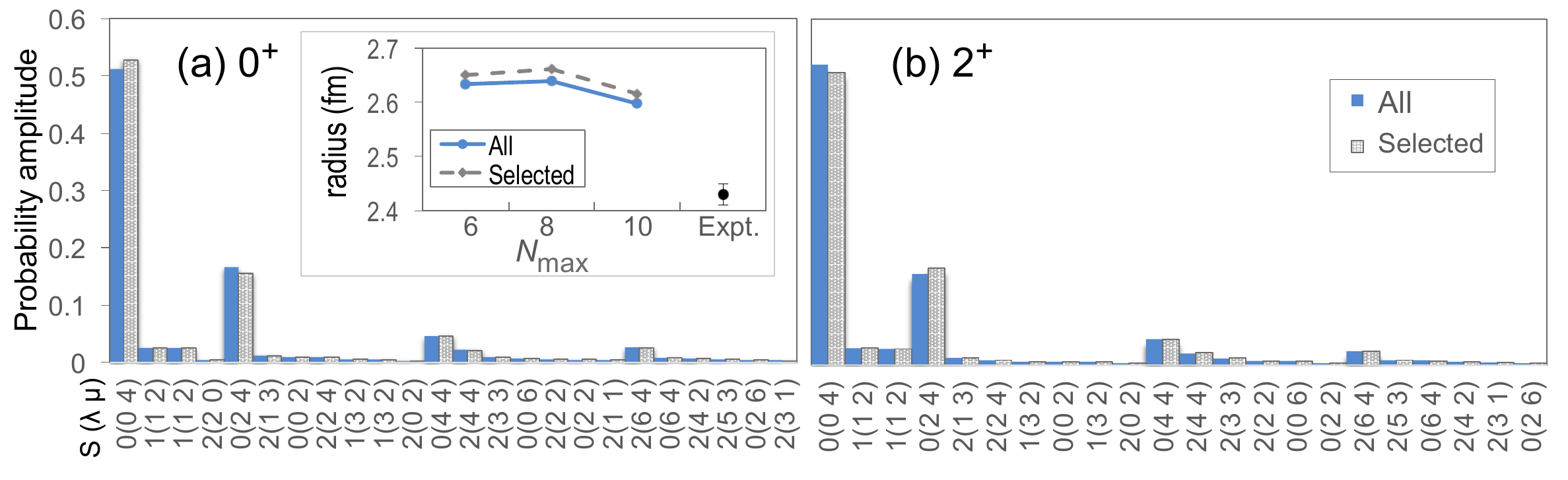} 
\caption{
 \SU{3} irreps  that make up (a) the ground state and (b) the first $2^+$  state of $^{12}$C, labeled by the  total intrinsic spin $S$ and \SU{3} labels $(\lambda\,\mu)$.  
Results are reported for {\it ab initio} SA-NCSM calculations with the full N3LO-EM interaction \cite{EntemM03} (solid blue) and its selected counterpart (dashed gray), for an \SU{3}-adapted basis in a complete model space of 8 HO major shells and $\hw= 15$ MeV. Insets:  matter rms radius of the $^{12}$C ground state as a function of the model space.
}
\label{su3_12C}      
\end{figure}

The predominance of a few shapes is neither sensitive to the type of the realistic interaction used, nor to the parameters of the basis, \hw~ and $N_{\rm max}$ \cite{DytrychLDRWRBB20}. Details such as contribution percentages slightly vary, but dominant features retain. Furthermore, even when the NN interaction is trimmed down by removing many SU(3)-symmetric components  that contribute less than a percent to the entire interaction, the results practically  coincide with the  corresponding \textit{ab initio} calculations that use the full interaction \cite{SargsyanLBDD20}.  As an illustrative example, we show that the \SU{3} content for both the ground state and the lowest $2^+$ state in $^{12}$C remains practically the same when the full N3LO-EM \cite{EntemM03} is used or its selected counterpart (Fig. \ref{su3_12C}). The corresponding  matter rms radius deviates only by 1\% when the selected interaction is used (Fig. \ref{su3_12C}a, inset), and such deviations typically decrease with larger model spaces (we note that in these calculations we neglect the three-nucleon forces that will reduce the deviation from the experimental value).  This study offers another remarkable outcome, namely, chiral potentials such as N3LO-EM, when expressed as a sum of SU(3)-symmetric components, exhibit a clear dominance of its  $(0\,0)$ component, which preserves deformation. In addition, we find that many of these components are negligible, which in turn makes the selection feasible.

Finally, the symplectic symmetry has been shown to ubiquitously arise from first principles  regardless of the NN interaction and the type of nucleus. It is then interesting to address the question whether all NN interactions used in such calculations possess close similarity to each other, or if they deviate in certain features that appear to be inconsequential to the emergence of the symmetry. To study this, we consider average energies per pairs (centroids) and correlations \cite{LauneyDD12}  between various realistic interactions  within the framework of spectral distribution theory \cite{French66,FrenchR71,ChangFT71,HechtD74,SviratchevaDV08,KotaH10} (Table \ref{correl}). The outcomes show that the average energy for an isoscalar (isovector) pair is the largest (smallest) for NNLO$_{\rm opt}$, as compared to other interactions, whereas all of the interactions correlate strongly both at the level of the interaction itself and as propagated in, e.g., $^{12}$C, with almost perfect correlation for the $T=0$ part of the NNLO$_{\rm sat}$ and NNLO$_{\rm opt}$. In general, two interactions with the same eigenvectors have a  correlation coefficient of $\zeta=1$. Hence, the correlation outcome  corroborates the above-mentioned  results, that is, the same orderly pattern is observed in the eigenvectors for all these interactions, while the emergence of the symmetry appears not to be sensitive to the differences in the average energy. 
\begin{table}
\caption{Average energy per pair $W_c$ (along the diagonal) and correlations $\zeta$ of selected phase-equivalent NN interactions (including the Coulomb interaction between protons), given as $(W_c^0,W_c^1)$ and $(\zeta^0,\zeta^1)$ for $(T=0,T=1)$ pairs, respectively. The NN part of two chiral potentials NNLO$_{\rm opt}$\cite{Ekstrom13}  and  NNLO$_{\rm sat}$ \cite{PhysRevC.91.051301} is considered, along with a comparison to the  JISP16 \cite{ShirokovMZVW07} based on the $J$-matrix inverse scattering method, for the two-nucleon system ($A=2$) and for a 12-particle system such as $^{12}$C ($A=12$, $T=0$). All interactions are reported for HO parameter $1/b \sim 0.6$ fm$^{-1}$.}
\label{correl}       % Give a unique label
% For LaTeX tables use
\begin{tabular}{llll|lll}
\hline\noalign{\smallskip}
& NNLO$_{\rm opt}$  &  NNLO$_{\rm sat}$  & JISP16 & NNLO$_{\rm opt}$  &  NNLO$_{\rm sat}$\\
& \multicolumn{3}{c}{$A=2$} & \multicolumn{2}{|c}{$A=12$, $T=0$} \\
\noalign{\smallskip}\hline\noalign{\smallskip}
NNLO$_{\rm opt}$  & $(-1.56, -0.19)$ &  &  \\
NNLO$_{\rm sat}$  & $(0.99,0.86)$ & $(-1.33, -0.54)$ & & $(0.92,0.92)$ & \\
JISP16  & $(0.88,0.84)$ & $(0.88,0.87)$ & $(-1.31, -0.51)$ & $(0.84,0.84)$ & $(0.88,0.88)$ \\
\noalign{\smallskip}\hline
\end{tabular}
\end{table}

In short, our findings show that nuclei below the calcium region, in their ground state as well as low-energy excitations, display relatively simple emergent physics that is collective in nature and tracks with an approximate symplectic symmetry heretofore gone unrecognized as emergent from the strong nuclear force. 
It is important to note that  no new dominant  shapes appear as we increase the model space, retaining the predominance of the single irrep, as shown in Ref. \cite{DytrychLDRWRBB20}. This has an important implication: complete SA-NCSM calculations are performed in smaller model-space sizes to identify the nonnegligible symplectic irreps, while the model space is then augmented by extending these irreps to high (otherwise inaccessible) HO major shells. Accessing these shells is vital to account for collective and spatially enhanced modes. As these modes play an important role in nuclear structure and reactions modeling, the present outcome is key to achieving {\it ab initio} predictions, e.g., for short-lived isotopes with deformed or cluster structure along various nucleosynthesis pathways, especially where experimental measurements are incomplete or not available.

\vspace{12pt}
\noindent
We acknowledge helpful discussions with D. J. Rowe, J. L. Wood, G. Rosensteel, J. P. Vary, P. Maris, C. W. Johnson, and D. Langr. This work was supported by the U.S. National Science Foundation (OIA-1738287, ACI -1713690, PHY-1913728), the Czech Science Foundation (16-16772S), and SURA, and in part by U.S. DOE (DE-FG02-93ER40756). This work benefitted from computing resources provided by Blue Waters,
LSU ({\tt www.hpc.lsu.edu}), the National Energy Research Scientific
Computing Center NERSC (a DOE Office of Science User Facility supported under Contract No. DE-AC02-05CH11231), and TACC's Frontera resource ({\tt www.tacc.utexas.edu}).

\vspace{12pt}
\noindent
{\it Author contributions.} --
All authors advanced the SA-NCSM framework, discussed the results and contributed to the final manuscript.
In addition, K.D.L. implemented the computer codes for correlations and analyzed the results;
T.D. implemented the suite of computer codes for the SA-NCSM and performed numerical simulations;
G.H.S. implemented the computer codes for selected interactions, performed and analyzed numerical simulations;
R.B.B. developed the theoretical formalism and implemented the computer codes for calculating response functions in the SA-NCSM, and performed the associated numerical simulations; and 
J.P.D. developed the theoretical formalism and computer package for \SU{3} coupling/recoupling coefficients.

\bibliographystyle{amsplainNoTitle}
\bibliography{launey_role_of_sp3r}
%\begin{thebibliography}{}
% and use \bibitem to create references.
%\bibitem{RefJ}
% Format for Journal Reference
%Author, Journal \textbf{Volume}, (year) page numbers
% Format for books
%\bibitem{RefB}
%Author, \textit{Book title} (Publisher, place year) page numbers
% etc
%\end{thebibliography}

\end{document}